\documentclass{emulateapj}
\usepackage{graphicx}
\usepackage{bm}
\usepackage{epsfig}
\usepackage{multirow}
\usepackage{color}

\shorttitle{Sky localization of compact binaries with gravitational waves}
\shortauthors{Nissanke et al.}

\begin{document}

\title{Localizing compact binary inspirals on the sky\\
using ground-based gravitational wave interferometers}

\author{Samaya Nissanke\altaffilmark{1,2}, Jonathan Sievers\altaffilmark{3}, Neal Dalal\altaffilmark{3,4}, Daniel Holz\altaffilmark{5,6}}

\altaffiltext{1}{JPL, California Institute of Technology, 4800 Oak
  Grove Drive, Pasadena, CA 91109, USA}

\altaffiltext{2}{Theoretical Astrophysics, California Institute of
  Technology, Pasadena, CA 91125, USA}

\altaffiltext{3}{CITA, University of Toronto, 60 St.~George St.,
  Toronto, ON, M5S 3H8, Canada}

\altaffiltext{4}{Department of Astronomy, University of Illinois, 1002
  W. Green St., Urbana, IL 61801, USA}

\altaffiltext{5}{Enrico Fermi Institute, Department of Physics, and
  Kavli Institute for Cosmological Physics, University of Chicago,
  Chicago, IL 60637, USA} 

\altaffiltext{6}{Theoretical Division, Los Alamos National Laboratory,
  Los Alamos, NM 87545, USA}

\begin{abstract}
The inspirals and mergers of compact binaries are among the most promising events for ground-based gravitational wave (GW)
observatories. The detection of electromagnetic (EM) signals from
these sources would provide complementary information to
the GW signal.  It is therefore important to determine the ability of GW detectors to 
localize compact binaries on the sky, so that they can be matched to
their EM counterparts.
We use Markov Chain Monte Carlo techniques
to study sky localization using networks of ground-based
interferometers. Using a coherent-network analysis, we find that the
Laser Interferometer Gravitational Wave Observatory (LIGO)--Virgo network
can localize 50\% of their detected neutron star binaries to better than
$50\,\mbox{deg}^2$ with 95$\%$ confidence region. The addition of the
Large Scale Cryogenic Gravitational Wave Telescope (LCGT) and LIGO-Australia improves this to $12\,\mbox{deg}^2$. Using a
more conservative coincident detection threshold, we find that 50\% of detected neutron
star binaries are localized to $13\,\mbox{deg}^2$ using the LIGO--Virgo
network, and to $3\,\mbox{deg}^2$ using the  LIGO--Virgo--LCGT--LIGO-Australia network. 
Our findings suggest that the
coordination of GW observatories and EM facilities offers great promise.
\end{abstract}


\section{Introduction}
\label{sec:intro}

The era of gravitational-wave (GW) astronomy is fast approaching. The advanced
versions of the Laser Interferometer Gravitational Wave
  Observatory (LIGO; ;~\citealt{Barish:1999}, \citealt{Sigg:2008}) and Virgo~(\citealt{Accadia:2011})
are expected to make their first detections within the coming
decade. Furthermore, construction has begun on the Large Scale
Cryogenic Gravitational Wave Telescope (LCGT;~\citealt{Kuroda:2010}), and an additional
advanced detector in Western Australia (referred to as LIGO-Australia
or LAu in this work) is
under serious consideration \citep{Barriga:2010,LIGOAustralia}. 
Inspiraling and merging compact-object binaries, composed of
neutron stars (NS) and/or stellar-mass black holes (BH),
are promising sources for these detectors.
For an advanced LIGO--Virgo network, predicted event rates
for NS--NS binaries range from
$0.4$ to $400$ year$^{-1}$ (with 40 being the ``realistic'' number
given in~\citep{Abadieetal:2010}) detectable to distances of several hundred Mpc, and approximately similar numbers apply for
NS--$10 M_\odot$ BH binaries with detectable distances $>1$ Gpc. 
Nearly face-on merging NS binaries are considered likely
progenitors of short hard $\gamma$-ray bursts
\citep[SHBs; e.g.,][]{eichler89}. \cite{Metzger:2010} suggest that NS
binary mergers, irrespective of their 
orientation, may produce radioactive decay powered transients with
absolute magnitude peak luminosities of $M_V = -15$ in the optical at $\sim$ day timescales. Thus, joint GW and electromagnetic (EM) observations can constrain
the physics of generic NS binary mergers, the central engine and outflows of
short GRBs, as well as cosmological parameters (see~\citealt{dalaletal};~\citealt[][henceforth N10]{Nissanke:2010}). GW events should, therefore, be
localized on the sky to sufficient accuracy to enable a match with their EM
counterparts. Henceforth, we use the terms ``sky localization'' to refer to the measured
sky position and its associated error uncertainty, while ``sky errors'' applies only
to area estimates.

For the typical NS binary inspiral event, whose duration is a few to tens of minutes in the detector frequency band, a single
interferometer has a broad antenna
pattern, and hence poor directional sensitivity. Two detectors
restrict the sky localization to a single ring. In a
network of three or more interferometers, relative arrival times of signals at each detector allow for the reconstruction of the
source's sky location to within a generally elliptical error
area of 1--100$\, \mbox{deg}^2$. For initial and advanced GW
detector networks, previous studies \citep{sylvestre, cavalieretal,
  blairetal,Fairhurst:2009, Klimenko:2011, Schutz:2011} have
explored sky errors for sources with both unmodeled (``burst'') and
modeled waveforms using analytically derived timing formulae and
Fisher-matrix methods. At high signal-to-noise ratio (SNR $>100$) 
such methods are
effective, providing lower bound error ellipses centered on the true
values of the parameters of interest. However, the
majority of sources for LIGO will be detected near threshold
($\mbox{SNR}\sim 8$). As discussed in N10 and~\cite{vallisneri08},
at ``low'' SNR ($\sim20$--$100$), 
parameter degeneracies can lead to quantitative and {\em qualitative}
errors in the Fisher-matrix approximation of the posterior
probability distribution functions (PDFs) for source parameters. Since
early GW detections will likely be low-SNR, a full treatment of the sky
localization will be critical when matching the GW event to an EM counterpart
both (1) to ensure high-probability regions are not falsely excluded from
consideration and (2) to minimize the number of false-positive transients that
must be dealt with. Later, should event rates reach tens per year as predicted,
the ease of finding counterparts will be directly related to how well GW
networks can localize events. In this paper we use Markov Chain Monte Carlo
(MCMC) techniques to map the full PDF of the sky location where an event may
have occurred.
Specifically, we examine sky localization for
non-spinning NS--NS populations 
for advanced GW detector networks, including LIGO (the two 4 km sites situated at Hanford and Livingston), Virgo, LCGT, and LIGO-Australia. For conciseness, we restrict our analysis here to
NS--NS binaries, results for NS--BH binaries being similar. For networks
comprising the initial versions of LIGO and Virgo detectors, several works (e.g., \citealt{rover07,vandersluys09,raymond09}) have addressed sky localization
using MCMC techniques for single NS--NS and spin-precessing
NS--BH systems. After the present article was submitted, a similar MCMC
study was performed by ~\cite{Aylott:2011}. They focus on the                   
addition of LIGO Australia, and find consistent results to ours. 

We follow similar techniques to those developed in~N10, which focuses on distance
determination using advanced GW detector networks for 
sources with known EM counterparts. In contrast to N10, 
here we instead ask how well a network of
ground-based GW detectors can determine the sky positions of previously
unknown compact binary
sources, a central question when planning complementary EM observations for counterparts.

This paper is organized as follows.  Section~\ref{sec:method} outlines the
extraction of sky position from the GWs emitted by inspiraling
binaries.
Section~\ref{sec:results} discusses our sky-localization results for
individual systems and populations of NS--NS binaries.
Section~\ref{sec:conclusion} presents our conclusions.


\section{Sky Localization Determination}
\label{sec:method}


Based on optimal matched filtering (\citealt{oppenheim83,finn92,cf94}), we extract the sky position ${\bf n}$ for
each NS--NS binary using knowledge of the expected GW waveform, where ${\bf n}
\equiv (\theta,\phi)$ is the vector pointing to a binary on the
sky (the waves therefore propagate to the Earth along $-\bf{ n}$). The
colatitude $\theta$ and longitude $\phi$ are related to the
declination $\delta$ and right ascension $\alpha$, by $\theta = \pi/2
- \delta$ and $\phi=\alpha-$GAST
respectively, where GAST is Greenwich Apparent Sidereal Time.  We use only the early inspiral portion of the
waveform, which for low mass systems provides most of the signal for advanced detectors \citep{Flanagan:1998} and is modeled accurately using post-Newtonian (PN) expansions in
general relativity. Specifically, we use the
non-spinning restricted 2PN waveform in the frequency domain for the
two GW polarizations $h_+$ and $h_{\times}$; see
Equations.~(12)--(14) in N10. The detector antenna functions depend on
${\bf n}$ and the binary's polarization angle. The overall
amplitude of the GW waveform encodes the source's orientation, sky location,
luminosity distance and redshifted chirp mass (see discussion in N10). For simplicity we assume that the inspiral
waveform ends abruptly prior to merger at the innermost stable
circular orbit. The time of flight from source at direction
${\bf n}$ to detector at location ${\bf r}$ involves the scalar
product ${\bf n}\cdot{\bf r}$, and differences in time of flight among
detectors in the network dominate sky localization.

In order to infer the sky position $(\cos\theta, \phi)$, we explicitly
map out the posterior PDF for all source parameters (including chirp
mass, orientation, etc.) given an observed data stream at a detector, using MCMC. The Metropolis--Hastings MCMC algorithm used is
based on a generic version of CosmoMC, described in \cite{lewis02}.
For simplicity, we assume Gaussian, stationary, and zero-mean noise that is independent and uncorrelated between detectors. We
take the anticipated noise sensitivity curve for a
single advanced LIGO detector, given in \cite{ligo_noise}, for
broadband tuning, to be representative of all our detectors, imposing
a low-frequency cut off of 10 Hz. Since we trigger detections off the {\it expected} and not the {\it observed} SNR, we do not incorporate selection effects into our
analysis. Therefore, we take prior
distributions in all source parameters to be flat over the region of sample space where
the binary is detectable at an expected network SNR = 3.5. The
expected network SNR is defined as the root sum square of the expected
individual detector SNRs. For each MCMC simulation, we derive solid angle areas over $(\cos
\theta, \phi)$ for 68$\%$, 95$\%$, and 99$\%$ confidence regions.

\subsection{Detected Binary Populations}
\label{sec:detbin}


The selection of our ``detected'' binaries is central
to the derivation of representative sky localization statistics for binary
populations (\citealt{Schutz:2011}). We simulate a million binaries out to $z = 1$,
assuming a constant comoving volume density
in a $\Lambda$CDM universe \citep{komatsu09}, and with random binary sky
positions and orientations. 
Each NS has a physical mass of $1.4 \,M_{\odot}$. 
We consider two plausible scenarios where each binary in the total
population is selected: (1) if its expected network SNR is greater
than a network threshold of 8.5 (Case I) and (2) if each
expected SNR at both LIGO Hanford and LIGO Livingston has a value of 6 or
higher and if the expected network SNR is above 12 (Case
II). We choose to threshold off the LIGO detectors, as opposed to any
other two detectors, only as a representative example out of all other possible coincident threshold scenarios. Case I uses a {\it network} thresholding criterion similar to that used in
N10 and \cite{Schutz:2011}. 
\section{Results and Discussion}
\label{sec:results}


\subsection{Individual Binaries}

We first examine sky localization for NS--NS binaries taken at random from our
Case I or Case II detected samples.  For each binary we assign a
unique noise realization to each detector, which we keep constant when
adding and subtracting detectors to a network. This enables a
meaningful comparison between the performance of different networks. We consider networks
comprising combinations of LIGO, Virgo, LAu, and LCGT.

\begin{figure}[t]
\centering 
\resizebox{\hsize}{!}{\includegraphics[width=0.95\textwidth]{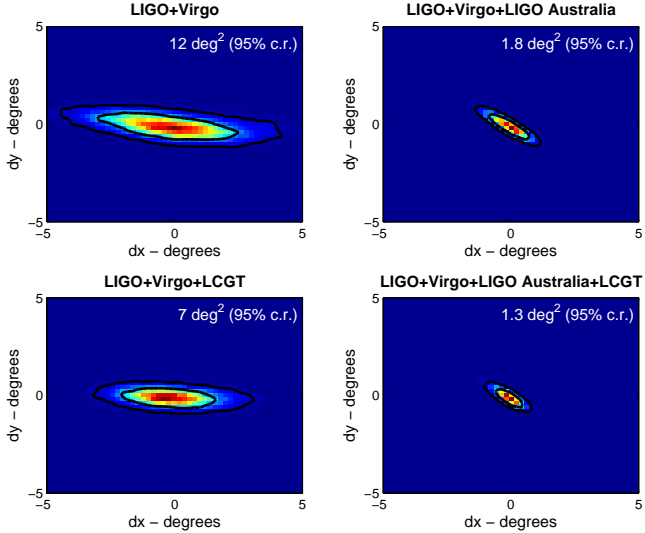}}
\caption{Sky localization for a high SNR binary under different network
  configurations, as labeled. The solid black curves indicate the 68\%
  and 95\% confidence regions (c.r.).  Additional detectors
  increase the network SNR and decrease the error ellipse, however
  note the significant localization enhancement provided by
  LAu in particular.  The origin (0,0) of each plot represents the
  source's true position, and the solid black lines denote the
  confidence regions.
}
\label{fig:highSNRellipse}
\end{figure}

In agreement with \cite{Fairhurst:2010} and \cite{Wen:2010}, we find
elliptically shaped errors for the majority of our examined NS--NS
binaries. This is unsurprising given that sky localization
reconstruction is dominated by differing GW arrival times at each 
detector rather than the direction-dependent antenna functions in the GW
amplitude.  For the three-element LIGO--Virgo network, 
sources located toward the detector plane produce
elongated error ellipses and have relatively poor angular resolution.
However, the error ellipse significantly decreases in
size as additional detectors are added to the network. The
inclusion of LAu is particularly favorable and breaks the {LIGO--Virgo(--LCGT)}
degenerate plane. As a representative example, Figure~\ref{fig:highSNRellipse}
shows a relatively high SNR GW signal for a NS--NS binary located at 180 Mpc with
an inclination angle of $\cos \iota = 0.7$ and a sky location of $(\cos
\theta = -0.3, \phi = 2.9)$. The expected SNRs at LIGO Hanford, LIGO
Livingston, Virgo, LAu, and LCGT are 6.7, 7.8, 12.4, 10.5, and 8.9,
respectively. The specific orientation and shape of the ellipse are
dependent upon the sky position and orientation, and noise realization.

\begin{figure}[t]
\centering 
\resizebox{\hsize}{!}{\includegraphics[width=0.95\textwidth]{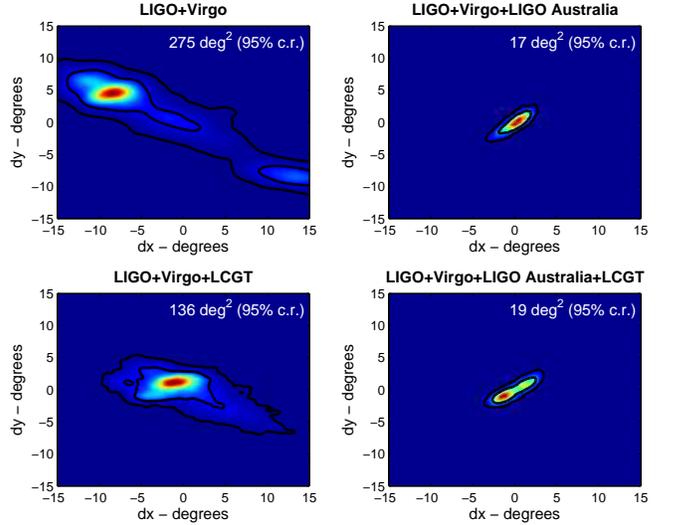}}
\caption{Sky localization for a low SNR binary. Contour levels are as
  in Figure~\ref{fig:highSNRellipse}.  In this example, the three-detector
  network finds multiple, widely separated islands of high likelihood,
  offset from the true location.  Additional detectors, however, can
  break underlying degeneracies. Furthermore, the binary's sky
  error at 95\% c.r. increases slightly with the five-detector network
  compared to the four-detector (LAu) network, because in this particular
  instance, the noise realization is unfavorable at LCGT.
}
\vspace{12pt}
\label{fig:lowSNRellipse}
\end{figure} 
However, for a handful of events near threshold, we find
that sky error areas are considerably non-ellipsoidal
and exhibit multimodal distributions, in particular with networks that
do not include LAu. As expected, their errors
are not centered on the source's true sky position and we find that
$5\%$ of our selected binaries have true positions that lie outside
the 95$\%$ confidence region. Analyses such
as~\cite{Fairhurst:2010}, \cite{Wen:2010}, and \cite{Schutz:2011} cannot reproduce such features due to
limiting assumptions implicit in timing and Fisher information methods.
Figure~\ref{fig:lowSNRellipse} shows an example of a low-SNR NS--NS
binary located at 567 Mpc with an
inclination angle of $\cos \iota = -0.93$ and a sky location of $(\cos
\theta = -0.36, \phi = 1.5)$. The expected SNRs at LIGO Hanford, LIGO
Livingston, Virgo, LAu, and LCGT are 5.4, 6.2, 3.1, 5.6, and 3.0, respectively.  

We find that error areas typically decrease by a factor of 4--5 by including LAu in
the network. For particularly weak signals, multimodal peaks and
statistical biases occur because multiple likelihood peaks for
different values of the GW waveform's amplitude become
indistinguishable from each other due to the uncertainties in the signal's
time of arrival at each detector.

\subsection{Detected Samples of Binaries}


We now examine cumulative distribution in sky errors for ensembles of GW NS--NS events
using different detector networks.  We randomly choose events from
our samples of detected NS--NS binaries using the
two different selection criteria detailed in Section~\ref{sec:detbin}.

\subsubsection{Case I: triggering on a Network of Detectors}

In Case I we set a total GW detector network threshold of 8.5, which
implies an {\it approximate} SNR threshold per detector
of $8.5/\sqrt{5}\sim3.8$ for each of the five detectors. As a first approximation, we estimate a detectable range of events within the maximum network
capability: each detector in the five-detector network has a ``weighted geometric average'' (optimal) range
of about $420\,{\rm Mpc} $ ($940\,{\rm Mpc}$) for NS--NS
events.\footnote{Similar to \cite{Abadieetal:2010}, we do not henceforth incorporate cosmological
  redshifts for our NS masses when estimating detectable
  ranges and rates. } The geometric average statistic used here is a weighted angular average over all sky positions and
orientations, which is a factor of $\sim 2.24$ smaller than the optimal
range for a face-on binary that is located directly above
the detector (see \citealt{Finn:1993}).

In order to obtain detection event rates, we follow the approach given
in~\cite{Abadieetal:2010} where detection ranges are derived by
thresholding off a single LIGO SNR of $8$. As the
noise in reality is non-Gaussian and non-stationary, \cite{Abadieetal:2010} use the
range for a single detector to represent the network of LIGO--Virgo detectors in order to achieve desired false alarm rates. For comparison purposes, we apply the same argument to our analysis: a five-detector network will have
an approximate threshold per detector
of $(8.5 \times \sqrt{3})/\sqrt{5}\sim6.6$, and hence a geometric average
(optimal) range of about $240\,{\rm Mpc}$ ($540\,{\rm Mpc}$) for
NS--NS events. \cite{Abadieetal:2010} provide a simple prescription to compute
the GW NS--NS detection rate using NS--NS
coalescence rates per galaxy, and the
number of galaxies accessible within a given GW reach. 
The NS--NS coalescence rate per galaxy is estimated either by extrapolating
from the observed sample of NS--NS binaries detected via pulsar
measurements, or by using population-synthesis methods. 
Using Table II with the associated {\it low}--{\it realistic}--{\it high}--{\it
  maximum} NS-NS coalescence rates per galaxy\footnote{\cite{Abadieetal:2010} assign rate estimates to one of four
  categories as detailed in their Section~IV; when rate PDFs are
  available, realistic refers to the mean of the PDF, low and high
  denote the $95\%$ pessimistic and optimistic confidence intervals,
  and maximum is the upper limit quoted in published literature.} and
Equations~(1) and (5) in \cite{Abadieetal:2010}, we define our {\it realistic} detection rate of $\sim 65$ NS--NS binaries per
year. Corresponding {\it low}, {\it high} and {\it maximum} detection
rates are, respectively, 0.7, $\sim$ 660 and $\sim$ 2650 NS--NS binaries per year seen
by the LIGO+Virgo+LAu+LCGT network. 

\begin{figure}
\centering 
\resizebox{\hsize}{!}{\includegraphics[width=0.95\textwidth]{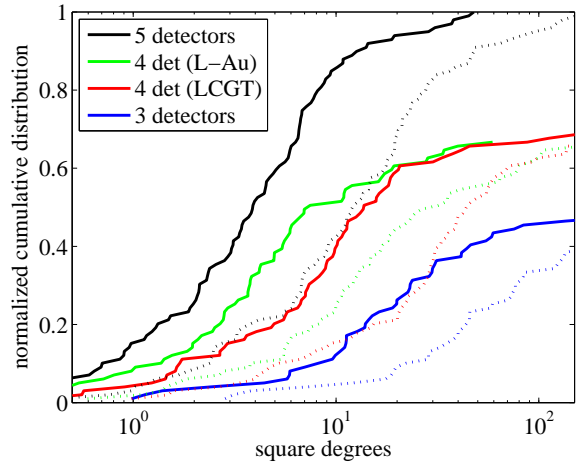}}
\caption{Normalized cumulative distributions as a function of the sky-error area (square degrees) of a sample of NS-NS binaries in Case I detection
scenario. Key: Solid/dotted lines denote 68\% and 95\%
c.r. respectively. Black: LIGO+Virgo+LAu+LCGT network,
green: LIGO+Virgo+LAu, red: LIGO+Virgo+LCGT, and
blue: LIGO+Virgo only.
}
\label{fig:cdcaseI}
\end{figure}

Turning to our results, we take a sample of $98$ NS-NS binaries detected by
the full five detector network, corresponding to an observation time
of $\sim$ 8 months based on our realistic rate estimates. In
Figure~\ref{fig:cdcaseI}, we show the normalized cumulative distribution of sky errors in square
degrees for our sample. We show distributions for subsets of systems detected by different networks 
which are normalized to the full sample illustrating the reduced number of detections. Table~\ref{tab:skylocerrors}
illustrates sky errors for $25\%$, $50\%$ and $75\%$ of NS-NS binaries from the
sample detected by a particular network. Among notable features: (1) the addition
of detectors to the network, in particular LAu, significantly reduces sky localization errors. We find that 50\% of all detectable NS-NSs are localized at 95\%
confidence region to within $10$--$20 \,\mbox{deg}^2$ with any
four or five detector network including LAu, and to within $110 \,\mbox{deg}^2 $ with
only the three LIGO+Virgo network. (2) The number of detected binaries doubles
as the numbers of detectors in a network increases from three to five.
 
\subsubsection{Case II: triggering on Individual LIGO Detectors}

For the Case II scenario we define detection using the more stringent
requirement of SNR $>$ 6 at each LIGO detector. Using an SNR of $10.4 \, \, (\sim 6 \times \sqrt{3})$ in each
LIGO detector (where the factor of $\sqrt{3}$ follows from the
\cite{Abadieetal:2010} correction for non-stationary and non-Gaussian
noise), we compute a weighted geometric average (optimal) range of 150 Mpc (340
Mpc). Invoking a similar argument as in Case I, we estimate a {\it realistic} detection rate of $\sim 17$ NS--NS binaries per
year. Corresponding low, high and maximum detection
rates are, respectively, 0.2, $\sim$ 165, and $\sim$ 660 NS--NS binaries per year.

\begin{figure}
\centering 
\resizebox{\hsize}{!}{\includegraphics[width=0.95\textwidth]{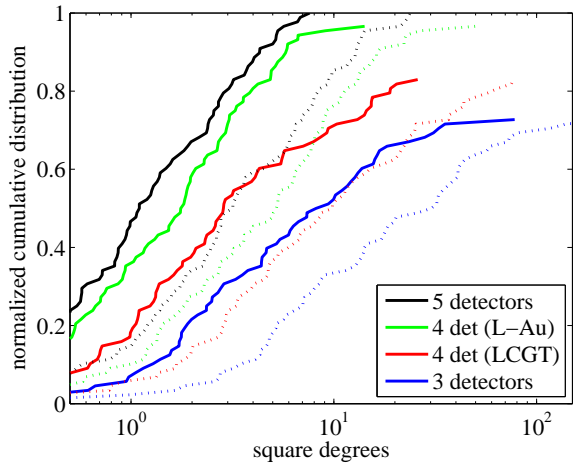}}
\caption{Normalized cumulative distributions of sky-error area (square degrees) for a sample of NS-NS binaries in Case II detection
scenario. The NS-NS detection rate in Case II is approximately four times
less than that in Case I. Key: Solid/dotted lines denote 68\% and 95\%
c.r. respectively. Black: LIGO+Virgo+LAu+LCGT network,
green: LIGO+Virgo+LAu, red: LIGO+Virgo+LCGT, and
blue: LIGO+Virgo only.}
\label{fig:cdcase2}
\end{figure}

Similar to Case I, we take a detected sample of $88$ NS--NS binaries seen by the
full LIGO-Virgo-LAu-LCGT network. Figure~\ref{fig:cdcase2} shows the normalized
cumulative distribution of our NS-NS binary sample as a function of sky
error in square degrees. We find that 50\% of NS--NSs are detected at 95\%
confidence level to within $5 \,\mbox{deg}^2$ with any network including LAu,
and $\sim 15$ $\,\mbox{deg}^2$ with only the LIGO--Virgo network. As before,
Table~\ref{tab:skylocerrors} shows the cumulative distribution of sky errors for
NS-NS binaries detected by different networks. Despite differences in our
analyses and astrophysical population models used, our results are consistent
with \cite{Fairhurst:2010} and \cite{Schutz:2011}. Once again, the inclusion of LAu substantially
improves localization errors and the binary detection rate rises with
more detectors.  In contrast to Case I, however, we find that the
number of detected binaries increases only by a factor of $1.3$ going
from a three- to five-detector network,
because both LIGOs must have SNRs $>$ 6. Using all five detectors, the
Case II scenario results in a factor of $\sim$ 5 fewer binaries than in the Case
I counterpart. We also find that the addition of a fourth or a fifth detector does
not improve sky coverage as significantly in Case II as in Case I.

\begin{deluxetable*}{llcccc}[t]
\tabletypesize{\scriptsize}
\tablewidth{16.0cm}
\tablecaption{Sky Errors in $\mathrm{deg}^2$ at 68$\%$/95$\%$/99$\%$
  Confidence Regions (c.r.) for NS--NS Binary Populations Detected Using Two Different Selection Criteria with Varying Networks. Percentages of detected binaries as a function
  of detector network are also indicated. }
\tablehead{
\colhead{Network} & \colhead{Fraction of binaries} & 
\multicolumn{1}{c}{LIGO+Virgo (LLV)} &
\multicolumn{1}{c}{LLV+LAu} &
\multicolumn{1}{c}{LLV+LCGT} &
\multicolumn{1}{c}{LLV+LAu+LCGT}  \\ \hline \\
\colhead{} & & 
\colhead{68$\%$/95$\%$/99$\%$} &
\colhead{68$\%$/95$\%$/99$\%$} &
\colhead{68$\%$/95$\%$/99$\%$} &
\colhead{68$\%$/95$\%$/99$\%$} \\
\colhead{} & & \colhead{in deg$^2$} &\colhead{in deg$^2$} &\colhead{in deg$^2$} &\colhead{in deg$^2$} 
}

\startdata

\multirow{6}{*}{Case I}& 25\% of binaries& 11/29/45&2/5/10&4/12/20&2/7/10 \\
&&&&& \\
& Median percentile&17/53/88&4/13/22&9/30/49&4/12/20\\
&&&&&  \\
& 75\% of binaries&30/124/200&7/30/53&15/47/82&7/22/39 \\
&&&&& \\ 
\hline \\
\multirow{6}{*}{Case II}& 25\% of binaries& 2/9/15&1/2/3&1/3/5&1/2/3 \\
&&&&& \\
&Median percentile &5/13/21&2/5/9&3/7/11&1/3/5\\
&&&&& \\
& 75\% of binaries&12/36/60&3/9/14&5/20/33&2/7/12 \\
&&&&& \\
\hline \hline \\
&&&&& \\
Case I& $\%$ of detected binaries& 49 & 66 & 69 & 100 \\
&&&&& \\
\hline \\
&&&&& \\
Case II& $\%$ of detected binaries& 73 & 97 & 83 & 100 \\
&&&&& 
\enddata

\label{tab:skylocerrors}
\end{deluxetable*}

\section{Conclusions}
\label{sec:conclusion}

This work examines sky localization for both individual systems and populations of
NS--NS binaries using different
networks of advanced GW
detectors. 
For the majority of
optimally oriented NS--NS binaries examined, we show good agreement of
our MCMC derived errors with the error
ellipses obtained from analytical timing accuracy and Fisher matrix
formulae. However, for a handful of SNR signals at threshold (in particular with the standard
geometrically degenerate LIGO--Virgo network), we show that sky error regions can be non-ellipsoidal
and non-contiguous, show multimodal distributions, and the best-fits can be
shifted away from their true values. Of particular relevance, our results show that the
inclusion of LIGO-Australia in a worldwide GW detector network improves
localization errors both for individual and populations of binaries up to a factor of $\sim$ 5,
reducing the appearance of multimodal islands. Finally, the number of detected binaries increases with the number
of detectors in a network.

A natural extension of this paper is to
include astrophysically realistic populations of NS--NS
and spin-precessing NS--BH binaries. Outstanding questions that are yet
to be addressed are the implications of our findings for observational EM
follow up. Our results show that measurements of astrophysical
populations of GW events result in error areas of $\sim 10\,
\mbox{deg}^2$. In the optical, cross-correlating localization error areas $> 10 \, \mbox{deg}^2$
with as complete as possible galaxy catalogs, such as the ``Local
Universe'' census proposed by \cite{kk09}, should aid EM search
strategies (\citealt{Kasliwal:2011}).  As recent observations
indicate, a small number ($\sim$ five at present) of SHBs appear to be located several tens of kpc away from
their host galaxies (e.g., \citealt{Berger:2010,Fong:2010}). Thus, search strategies will need to include
the possibility of such effects when looking for transients of
NS binary mergers. Moreover, time-domain surveys provide an estimate
of the false-positive rate of dynamic galactic (foreground) and extragalactic
(background) transients. Currently operating and future optical and radio EM facilities
are thus capable of preparing for and performing follow-up of GW events. 

\section{Acknowledgements}

We are grateful to Josh Bloom, Scott Hughes, Sterl Phinney, Bangalore
Sathyaprakash and Michele Vallisneri for suggestions in the early development of the work.  
We thank Chad Galley, Mansi Kasliwal, Ilya Mandel and Eran Ofek for careful reading of the
manuscript, and Yanbei Chen, Curt Cutler, Wen-Fai Fong, Chris Hirata, Tom Prince, Bernard Schutz, Kip
Thorne, Linqing Wen and Stan Whitcomb for helpful discussions. 
Simulations were performed using the Sunnyvale cluster at CITA, which is funded
by NSERC and CIAR.  
SMN's research was
carried out at the Jet Propulsion Laboratory, California Institute of
Technology under a contract with National Aeronautics and Space Administration.


\end{document}